\documentclass[a4paper]{jpconf}

\usepackage{graphicx}

\begin{document}

\title{Fluctuations and NA49}

\author{Marek Gazdzicki}

\address{University of Frankfurt, Germany and\\
Swietokrzyska Academy, Kielce, Poland}

\ead{Marek.Gazdzicki@cern.ch}

\begin{abstract}
A brief history of the study of fluctuations in high
energy nuclear collisions at the CERN SPS performed by  NA49
is presented.
The ideas and the corresponding experimental data on
fluctuations are discussed
from the point of view of their sensitivity to the 
onset of deconfinement.

\end{abstract}.

\section{Introduction}
What are the phases of strongly interacting matter? 
This question and 
in particular, the hypothesis that at high energy densities
the matter is in the form of quark-gluon plasma (QGP)~\cite{qgp} rather
than a gas of hadrons  motivated the first stage of the broad
experimental program of study of ultra-relativistic nucleus--nucleus
collisions \cite{exp}.
The results on the energy dependence of 
the  mean hadron multiplicities and the
shape of the transverse mass spectra obtained by the NA49
experiment at the CERN SPS indicate 
that the properties of the created matter
change rapidly in the region of low SPS energies \cite{onset}
($\sqrt{s}_{NN} \approx$ 6--12 GeV).
The observed effects confirm predictions of the transition from
hadron gas to QGP \cite{gago} and thus indicate that in fact
a new form of strongly interacting matter exists in nature at
sufficiently high energy densities. 

Do results on energy dependence of event-by-event fluctuations 
confirm the conjecture of the onset of deconfinement at low
SPS energies? 
This report summarizes the current status of the NA49 data on this
subject.

The paper starts with a brief historical review  of the event-by-event
physics in NA49 (Section~2). The experimental status of the
basic ideas relating the onset of deconfinement
and the fluctuations in nuclear collisions  are 
presented and discussed in Section 3.
A  summary given in Section 4 closes this report.

\section{A brief history of fluctuations in NA49}

The NA49 large acceptance spectrometer \cite{nim} was constructed with
the aim to study event-by-event fluctuations.
In a letter of intend \cite{loi} it was argued:
"the focus on an analysis of macroscopic 
and microscopic observables, referring to the
{\it individual events}, stems from our expectation that
the phase transition, due to the statistical fluctuations in the 
collision dynamics, may not uniformly occur in the {\it average}
central Pb+Pb collision". 
The first  data on central Pb+Pb collisions at 158$A$ GeV were taken
during the  NA49 running period 1994-1996 and soon after the
pioneering analysis of event-by-event fluctuations started. It appeared to be more
difficult than originally expected. In particular the problems were
caused by: 
\begin{itemize}
\item
a need to control the experimental biases on event-by-event basis,
\item
a difficulty to remove the influence of trivial fluctuations in  collision
geometry,
\item 
the absence of quantitative models predicting event-by-event fluctuations
in  case of the deconfinement phase transition.
\end{itemize}
Numerous measures of fluctuations were proposed, 
see e.g. Refs.~\cite{measures}, and their
advantages and disadvantages were vividly discussed.
Clearly a choice of the measure depends on the physics goals. It should
maximize  the sensitivity to the effects which we are looking for and
minimize the  sensitivity  to the biasing contributions (experimental and
physical).

\begin{figure}[h]
\begin{minipage}{18pc}
\includegraphics[width=14pc]{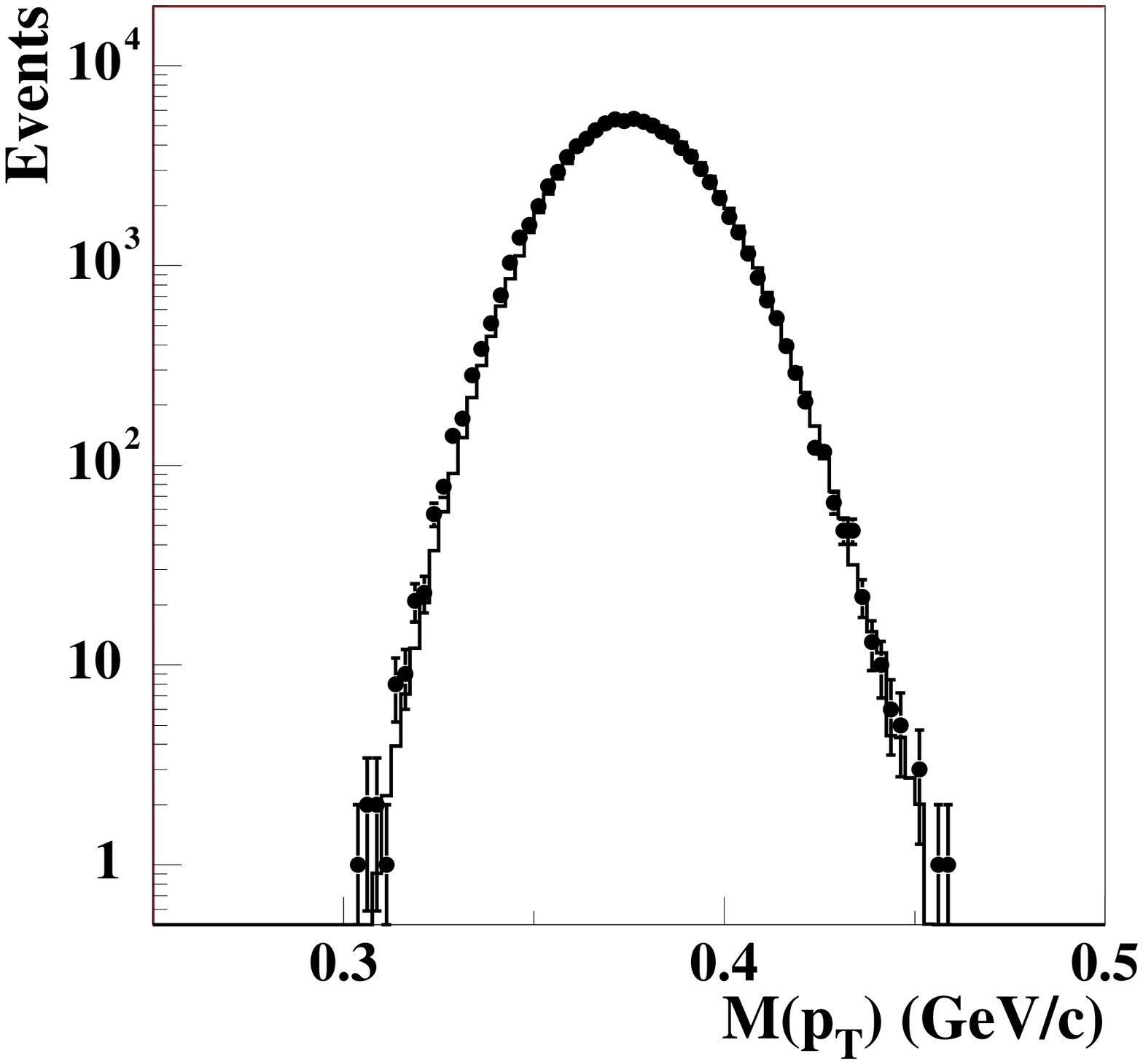}
\caption{\label{fig_mpt}
Event-by-event distribution of the mean transverse
momentum M(p$_T$) of accepted particles in the event
(points) for central Pb+Pb collisions at 158$A$ GeV. 
The solid line shows the M(p$_T$) distribution
for "mixed events" (histogram).}
\end{minipage}\hspace{1pc}%
\begin{minipage}{18pc}
\includegraphics[width=13pc,angle=270]{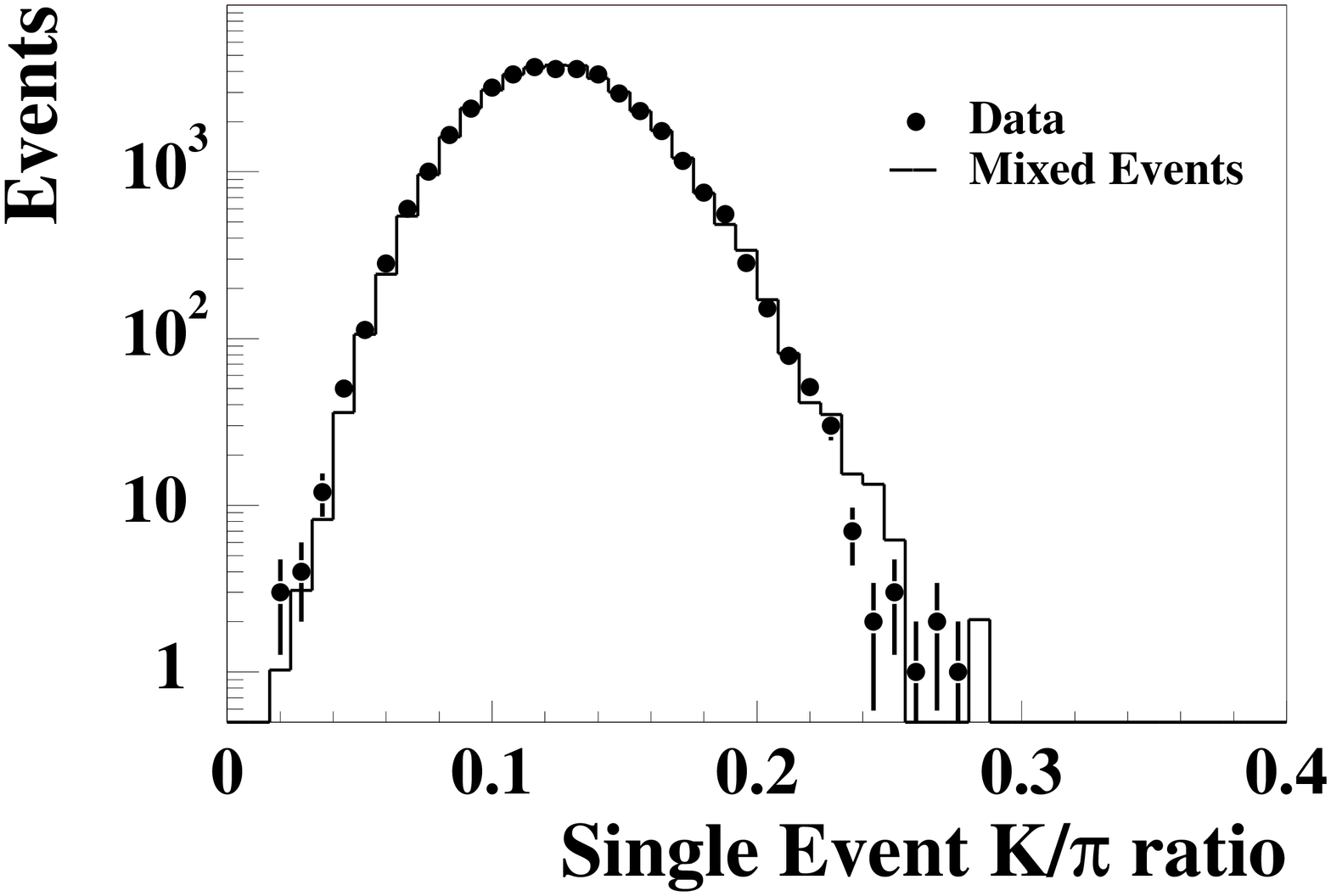}
\caption{\label{fig_kpi}
Distribution of the event-by-event kaon to pion ratio
estimated using a maximum likelihood method (points)
for central Pb+Pb collisions at 158$A$ GeV.
As a reference, the same procedure was applied to a "mixed
event" sample (histogram).}
\end{minipage} 
\end{figure}

Due to all these problems
the first results on transverse momentum fluctuations
\cite{Appelshauser:1999ft}
and the fluctuations of the kaon to pion ratio
\cite{Afanasev:2000fu}
in central Pb+Pb collisions were published only in 1999 and
2000, respectively.
The measured fluctuations (see Figs. \ref{fig_mpt} and \ref{fig_kpi}) 
are approximately
reproduced by a "mixed event" model 
in which uncorrelated production of hadrons is assumed.
They clearly demonstrate that  possible fluctuations of the early
stage energy density within the analyzed event sample, if existent, 
do not lead to significant changes of the event
properties. All collisions are "similar".  

\begin{figure}[h]
\begin{center}
\includegraphics[width=20pc]{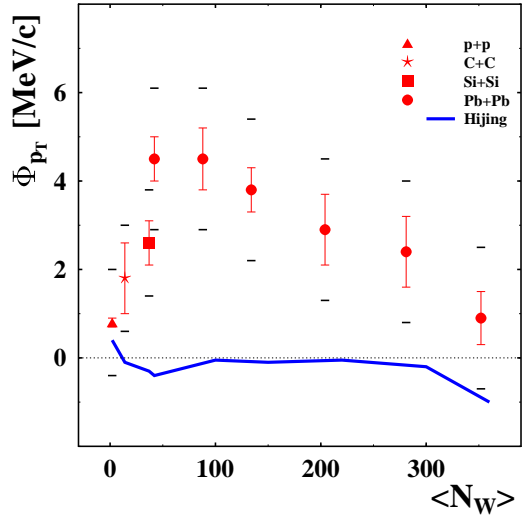}\hspace{2pc}%
\end{center}
\caption{\label{mpt}
The system size dependence of transverse momentum fluctuations
at 158$A$ GeV. The  points correspond to the
NA49 experimental data \protect\cite{kasia}, whereas the solid line to the
predictions of the string-hadronic model Hijing \protect\cite{hijing}.  
}
\end{figure}

In 1997 and 1998 the data on C+C and Si+Si collisions at 158$A$
and 40$A$ GeV were registered.  Together with the data
on all inelastic Pb+Pb interactions at these energies, they allow
to study system size dependence of event-by-event fluctuations.
In particular, 
the transverse momentum \cite{kasia} and multiplicity
\cite{maciek} fluctuations were analyzed.
The data on mean $p_T$  fluctuations, see Fig.~\ref{mpt},
indicate an increase of fluctuations for 
C+C, Si+Si and semi-peripheral Pb+Pb collisions.
A similar effect is also observed  at RHIC energies \cite{rhic}.
It is still
lacking a well established interpretation.

The hypothesis that the onset of deconfinement is located at
low SPS energies \cite{gago} motivated the energy scan program of the
NA49 Collaboration \cite{add1}. Within this program data on central
Pb+Pb collisions at 20$A$, 30$A$, 40$A$ and 80$A$ GeV were registered
in runs in 1999, 2000 and 2002.
The predicted \cite{gago} rapid changes in the energy dependence of hadron properties
averaged over central Pb+Pb collisions, like pion and kaon
mean multiplicities and transverse mass spectra, were observed \cite{onset}. 
The search for the expected anomalous behavior of other quantities
is in progress.  
Among them event-by-event fluctuations are of particular importance.
This is because they are expected to be sensitive 
to the physics properties of the system
which are not accessible in the
study of the mean quantities.
In the following section we review current status of the ideas and
the experimental results on the energy dependence of electric charge,
multiplicity and strangeness fluctuations in the domain of the onset
of deconfinement.

\section{Fluctuations and the onset of deconfinement}

\subsection{Net-electric charge fluctuations}
\begin{figure}[h]
\includegraphics[width=36pc]{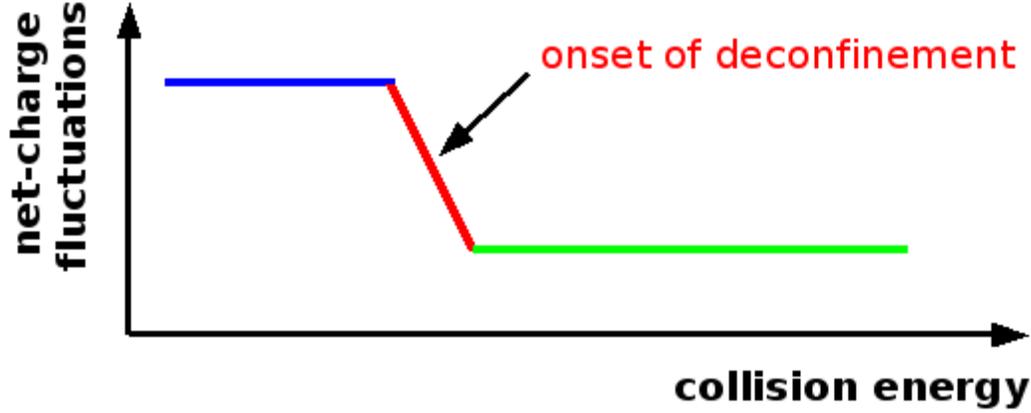}\hspace{2pc}%
\caption{\label{nc}
A naive expectation for  the energy dependence of the
net-electric charge fluctuations in the region of the
onset of deconfinement.}
\end{figure}

A simple estimate of the net electric charge fluctuations show
that they are much smaller in a Quark-Gluon Plasma than in a hadron
gas \cite{jeon}. 
This difference is  caused by smaller charge units in the QGP (fractional
charges of quarks) than in the hadron gas. 
Thus a decrease of the net-charge fluctuations is expected when
the collision energy crosses the threshold for the deconfinement
phase transition. 
A schematic sketch of this naive prediction is presented in Fig.~\ref{nc}. 

\begin{figure}[h]
\includegraphics[width=36pc]{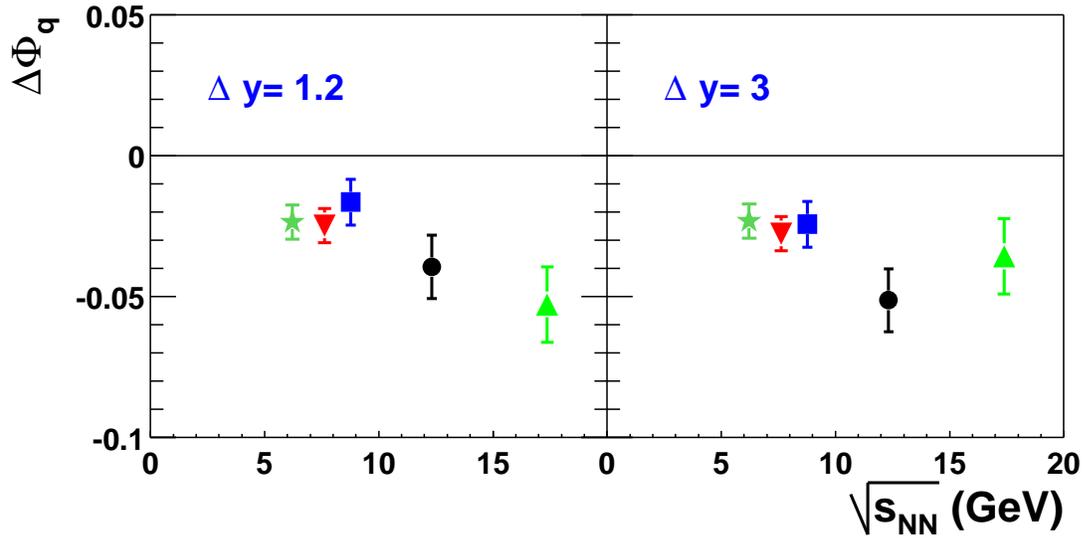}\hspace{2pc}%
\caption{\label{nc_data}
The energy dependence of the net-charge fluctuations measured by
$\Delta \Phi_Q$ \protect\cite{Zaranek:2001di} in central Pb+Pb collisions
for two different rapidity intervals $\Delta y = 1.2$ (left) and
$\Delta y = 3$ (right). For more details see the original publication
\protect\cite{Alt:2004ir}. }
\end{figure}

\begin{figure}[h]
\begin{center}
\includegraphics[width=20pc]{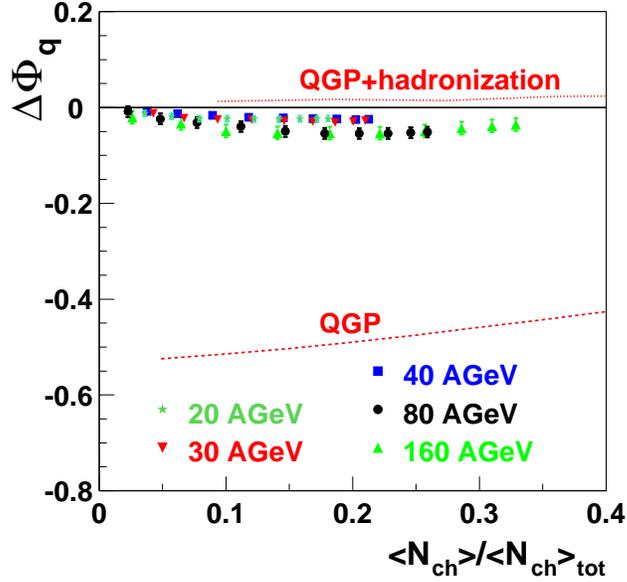}\hspace{2pc}%
\end{center}
\caption{\label{nc_data1}
The dependence of $\Delta \Phi_q$ on the fraction of accepted
particles in central Pb+Pb collisions at 20-158$A$ GeV. The prediction for
the ideal QGP is indicated by the dashed curve (QGP), whereas
the prediction for the QGP including hadronization and
resonance decay is shown by the dotted curve (QGP+hadronization). 
For more details see the original publication
\protect\cite{Alt:2004ir}. }
\end{figure}

The experimental results on central Pb+Pb collisions
at 20$A$, 30$A$, 40$A$, 80$A$ and 158$A$ GeV shown
in Fig.~\ref{nc_data} suggest
only a very weak, if any, dependence on energy.
They are close to those expected for a gas of pions
correlated only by global charge conservation 
($\Delta \Phi_q = 0$) and significantly above the
naive prediction for the QGP $\Delta \Phi_Q \approx -0.5$
\cite{Zaranek:2001di}.
These results do not necessarily exclude reduced
fluctuations in the QGP because these can be masked by contributions
from resonance decays.
In fact a simple model of the QGP hadronization and resonance decay
can quantitatively explain the magnitude of the measured fluctuations, see 
Fig.~\ref{nc_data1}. 

\begin{figure}[h]
\includegraphics[width=36pc]{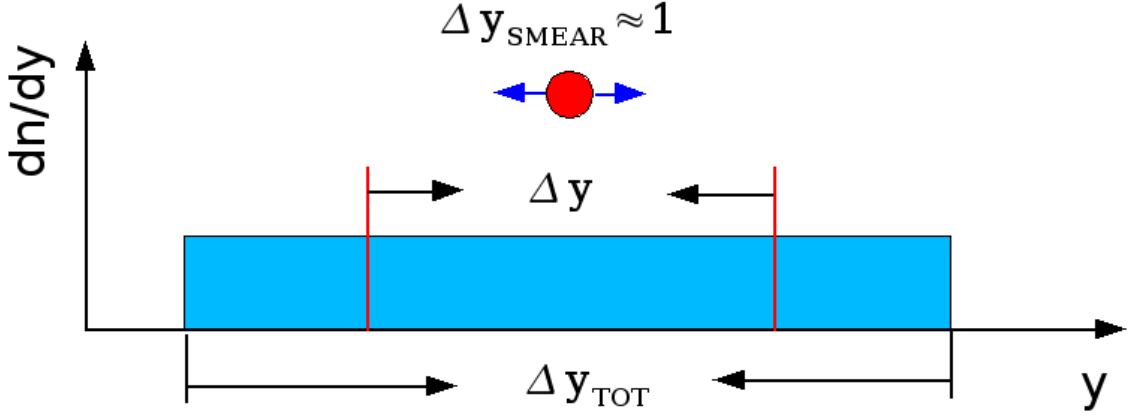}\hspace{2pc}%
\caption{\label{sketch_smear}
A sketch illustrating different rapidity scales
important in
the study of the net-charge fluctuations as a possible signal
of QGP. 
$\Delta {\rm y}$ is the acceptance window in rapidity, 
$\Delta {\rm y_{SMEAR}}$ is a typical distance in rapidity of 
decay products of resonances and 
$\Delta {\rm y_{TOT}}$ is the width of the rapidity distribution
of all particles. 
The observation of the predicted suppression of the
net-charge fluctuations in QGP requires:
$\Delta {\rm y_{SMEAR}} < \Delta {\rm y} < \Delta {\rm y_{TOT}}$.
This condition is not fulfilled at the SPS energies where
$\Delta {\rm y_{SMEAR}} \approx \Delta {\rm y} \approx \Delta {\rm y_{TOT}} 
\approx 1$.
 }
\end{figure}

The influence of resonance decays on charge fluctuations depends on the
size of the rapidity interval, $\Delta {\rm y}$, in which fluctuations are calculated.
If $\Delta {\rm y}$ is much bigger than the typical distance in rapidity
of the daughter particles, $\Delta {\rm y_{SMEAR}}$, the charge within the interval
will not be changed by the decay and therefore the charge fluctuations
should not be affected. On the other hand, if $\Delta {\rm y}$ is small, a large
fraction of daughter particles will leave the interval and the initial net-charge
will be significantly changed. The mean rapidity difference of two pions
originating from decays of $\rho(770)$ meson is approximately one unit
of rapidity. Therefore in order to minimize the decay effect $\Delta {\rm y}$
should be much larger than 1.
However, this constraint is difficult to fulfill at SPS and
lower energies because the width of the rapidity distribution of all
produced particles, $\Delta {\rm y_{TOT}}$ is not much broader than 1.
Hence a rapidity interval which is large enough to be unaffected by the 
resonance decays would contain almost all particles
produced in a collision. The net-charge in this interval would then reflect
the number of participant protons and the fluctuations would be
determined by fluctuations of the collision centrality and not the particle
production mechanism.  
A  sketch which illustrates these considerations is shown in
Fig.~\ref{sketch_smear}.
At  the SPS and AGS energies 
$\Delta y \approx \Delta y_{SMEAR} \approx  \Delta y_{TOT} \approx 1 $
and consequently
the net-charge fluctuations are not sensitive to the initial QGP
fluctuations.
Hence their energy dependence may be
not affected by  the onset of deconfinement.

\subsection{Multiplicity fluctuations}
\begin{figure}[h]
\includegraphics[width=36pc]{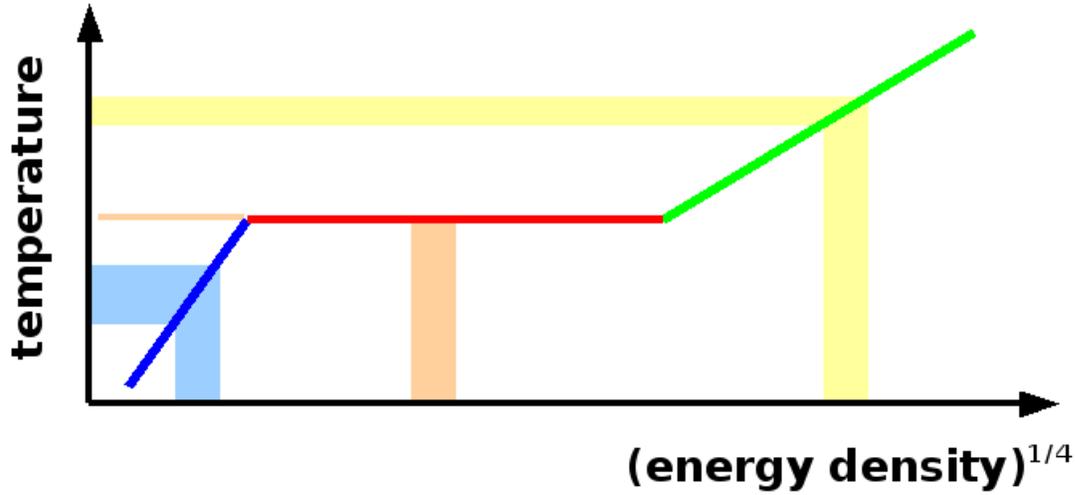}\hspace{2pc}%
\caption{\label{sketch_temp}
A sketch illustrating how the early stage energy density
fluctuations lead to temperature fluctuations which
depend on the equation of state.
The solid line shows equation of state with the 1st
order phase transition. The width of the vertical bands indicates
early stage energy density fluctuations for three different
collision energies, whereas the width of the horizontal
bands gives the corresponding temperature fluctuations.   
}
\end{figure}

It was suggested that
properly filtered multiplicity fluctuations should
be sensitive to the equation of state at the early stage
of the collision and thus to its changes in the 
deconfinement phase transition region \cite{mfluct}.
This idea follows from the expectation that the early stage
energy density changes from collision to collision and thus it 
causes fluctuations of the thermodynamical parameters of the
matter like temperature, pressure, entropy and strangeness. 
Obviously the relation between energy density fluctuations and
fluctuations of the other thermodynamical parameters depends
on the equation of state. 
This opens a possibility to study equation of state via
properly defined analysis of event-by-event fluctuations.
This basic idea is sketched in Fig.~\ref{sketch_temp} for the
case of the temperature fluctuations in the phase transition
domain.

As the entropy is closely related to the particle
multiplicity, and it is expected to be approximately conserved
during the evolution of the matter created at the early
stage,    the fluctuations of entropy are of primary
interest.
Based on the second principle of thermodynamics 
it can be  shown \cite{mfluct}
that the relative entropy fluctuations are given by:
\begin{equation}
\frac {(\delta S)^2} {S^2} = 
(1+ \frac {p} {\epsilon})^{-2} \frac {(\delta E)^2} {E^2}, 
\end{equation}
where $S$ and $E$ denote entropy and energy, respectively,
$p$ is pressure and $\epsilon$ energy density.
For the experimental study the measure of multiplicity
fluctuations, $R_e$ was proposed \cite{mfluct}:  
\begin{equation}
R_e  = \frac {(\delta \bar n)^2/\bar n^2} {(\delta \bar E)^2/ \bar E^2},
\end{equation}
where
$\delta \bar n$ and $\delta \bar E$ are dynamical fluctuations of
particle multiplicity and energy, respectively.
They are related to the variance of particle multiplicity $V(n)$ as:
\begin{equation}
V(n) = 
(\delta \bar  n)^2 + \langle (\delta  n)^2 \rangle ,
\end{equation}  
where $\langle (\delta  n)^2 \rangle$ are dynamically
averaged statistical fluctuations which can be measured using
the sub-event method \cite{volo}.
The measure $R_e$ is directly related to the early stage equation of
state \cite{mfluct}:
\begin{equation}
R_e = (1+ \frac {p} {\epsilon})^{-2},
\end{equation}
and therefore it should be sensitive to the onset of deconfinement. 
As it is defined as a ratio of relative multiplicity and
energy fluctuations it is insensitive to the magnitude of the 
early stage energy density fluctuations.
One should underline that the method assumes that the early
stage entropy fluctuations are proportional to the dynamical
multiplicity fluctuations, but it does not require knowledge of
the value of the proportionality factor. 
This is a clear advantage in regard to the methods of studing the equation
of state based on the analysis of the average quantities.

\begin{figure}[h]
\begin{minipage}{18pc}
\includegraphics[width=18pc]{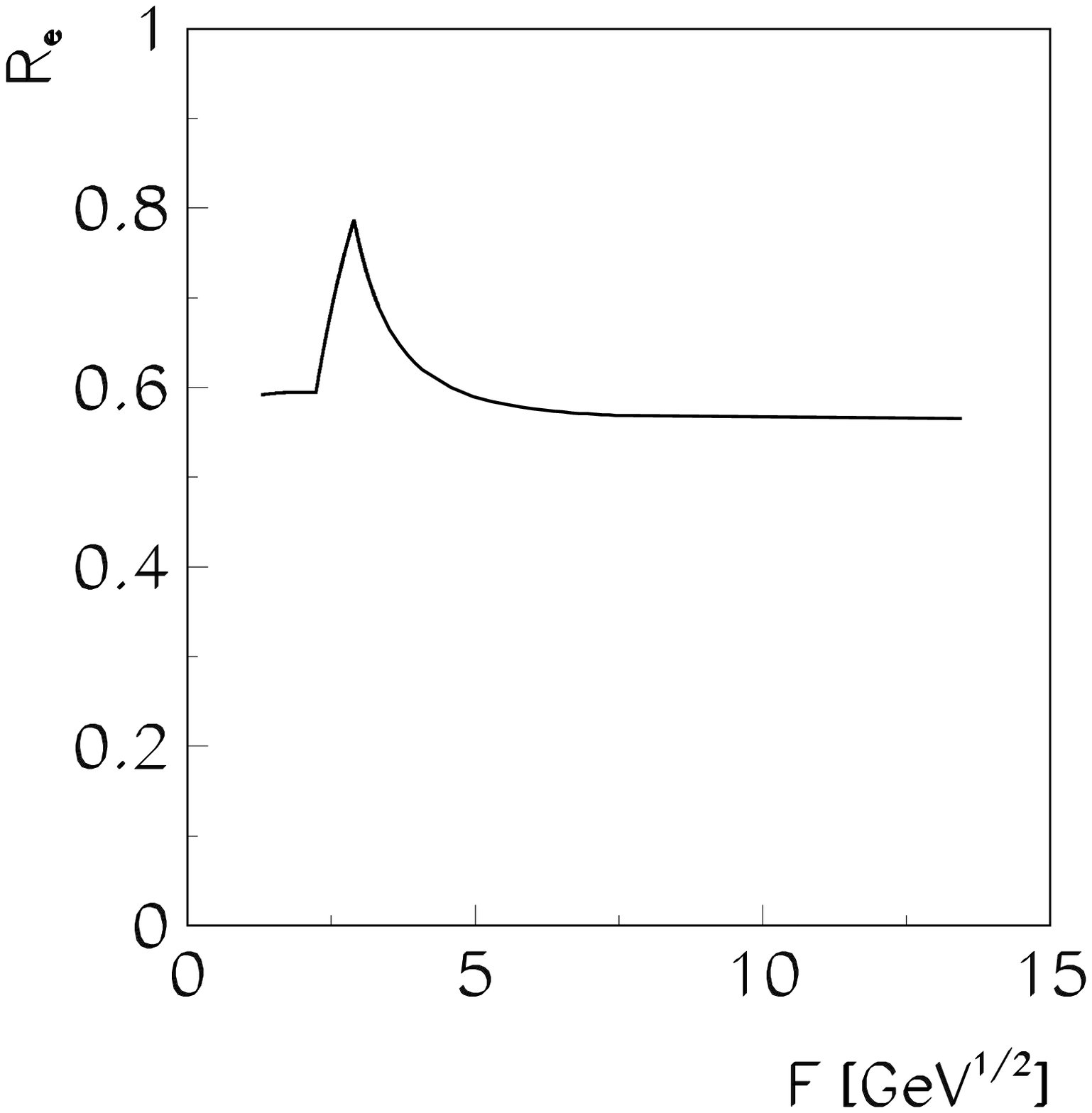}\hspace{1pc}
\caption{\label{shark}
The energy dependence ($F \equiv (\sqrt{s} - 2 m)^{3/4}/s^{1/8}$) 
of the properly filtered multiplicity fluctuations \protect\cite{mfluct},
$R_e$, calculated with SMES \protect\cite{gago,mfluct}.
The 'shark fin' structure at the low SPS energies is caused by
the onset of deconfinement. 
}
\end{minipage}\hspace{1pc}%
\begin{minipage}{18pc}
\vspace{3pc}\hspace{-1pc}\includegraphics[width=20pc]{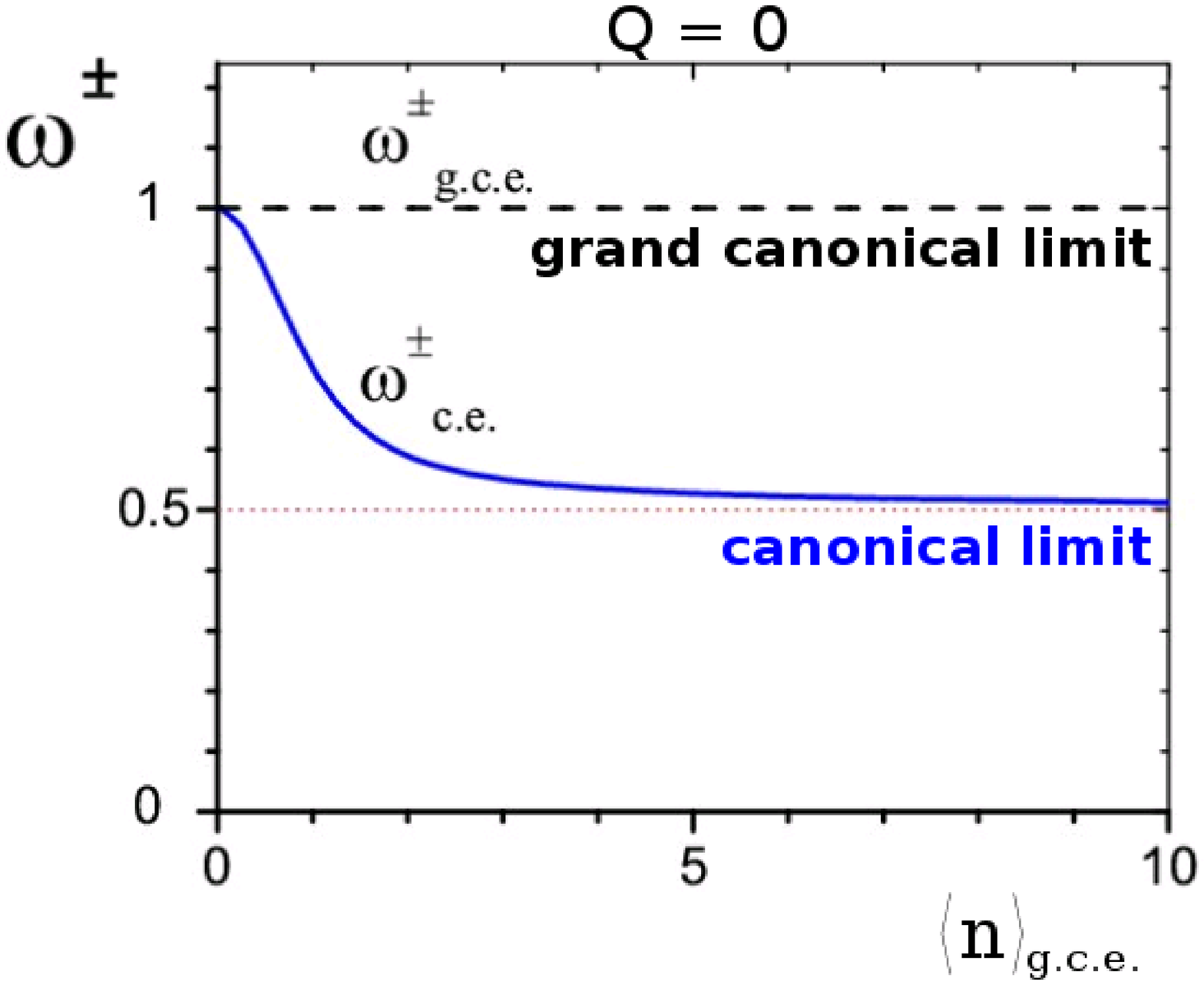}
\vspace{0.5pc}\caption{\label{w}
The scaled variance for positively (+) and negatively (-) charged
particles calculated within canonical and grand canonical
ensembles as a function of the mean multiplicity
\protect\cite{m1}.
The results corresponds to  a system with net charge equal to zero. 
}
\end{minipage}%
\end{figure}

The energy dependence of $R_e$ calculated with the Statistical Model
of the Early Stage (SMES)~\cite{gago} is presented in 
Fig.~\ref{shark} \cite{mfluct}.
It shows a characteristic 'shark fin' structure caused by  large
fluctuations in the mixed phase region.

This simple idea motivated further theoretical and experimental effort
directed toward understanding of multiplicity fluctuations in 
nucleus-nucleus collisions at high energies.

The effect of conservation laws on the multiplicity fluctuations 
were studied within canonical and micro-canonical ensembles.
This led to an unexpected observation \cite{m1} that the
scaled variance of the multiplicity distribution,
$w = V(n)/\langle n \rangle$, is different in various statistical
ensembles (grand canonical, canonical and micro-canonical)  
even in the thermodynamical limit, $V \rightarrow \infty$.
The simplest example is presented in Fig.~\ref{w}, where
the scaled variance for positively and negatively charged
particles calculated within grand canonical and canonical
ensembles is plotted as a function of mean particle multiplicity
\cite{m1}.
A detailed study of multiplicities fluctuations
in different statistical ensembles is in progress \cite{m2,m3}. 

An important contribution to multiplicity fluctuations may
come from the fluctuations in the collision geometry. 
Clearly an interaction with a small number of participating nucleons
will result in much lower particle multiplicity than
a central collision in which almost all nucleons are involved in the
reaction process.
\begin{figure}[h]
\hspace{8pc}\includegraphics[width=20pc]{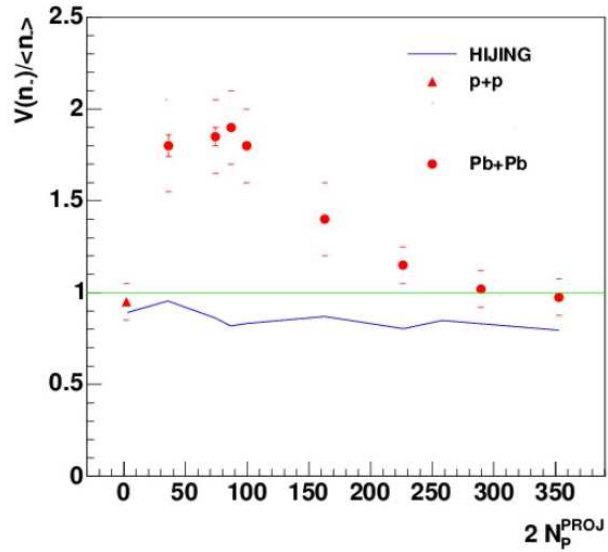}
\caption{\label{maciek}
The scaled variance of the multiplicity distribution for the negatively  charged
hadrons produced in Pb+Pb collisions (circles) and p+p interactions
(triangle)
at 158$A$ GeV as a function of the number of participants nucleons
from the projectile \protect\cite{maciek}.
The solid line indicates  predictions of a string-hadronic
model Hijing \protect\cite{hijing}. 
}
\end{figure}
Thus in order to extract the  dynamical effects a precise control
of the geometrical fluctuations is needed.
In NA49 this is achieved by selecting events with fixed number 
of projectile participants, $N^{PROJ}_P$, measured by 
the forward VETO calorimeter \cite{veto}.
The presented results are corrected for the resolution of the
VETO calorimeter and  the finite bin size in the measured 
forward energy.
In contrary to the naive expectations~\cite{hijing} 
it was found~\cite{maciek} that
the scaled variance of the multiplicity distribution increases
with a decreasing centrality of the Pb+Pb collisions at 158$A$ GeV.
The preliminary results\footnote{
Note that an error was found in the analysis of multiplicity
fluctuations in C+C and Si+Si collisions at 158$A$~GeV, and thus the
corresponding preliminary results shown in Refs. \cite{maciek,marek} 
should not be used any more.}
are shown in  Fig.~\ref{maciek}.
The observed non-trivial centrality dependence focused the
efforts on its tests and understanding and consequently 
the analysis of the energy dependence of multiplicity
fluctuations was delayed.

\subsection{Strangeness fluctuations}

Similar to entropy (multiplicity) fluctuations, 
strangeness fluctuations are also sensitive to the
early stage energy density fluctuations and via these
to the equation of state at the early stage of the
collisions \cite{sfluct}.
\begin{figure}[h]
\begin{minipage}{18pc}
\vspace{1pc}\includegraphics[width=17pc]{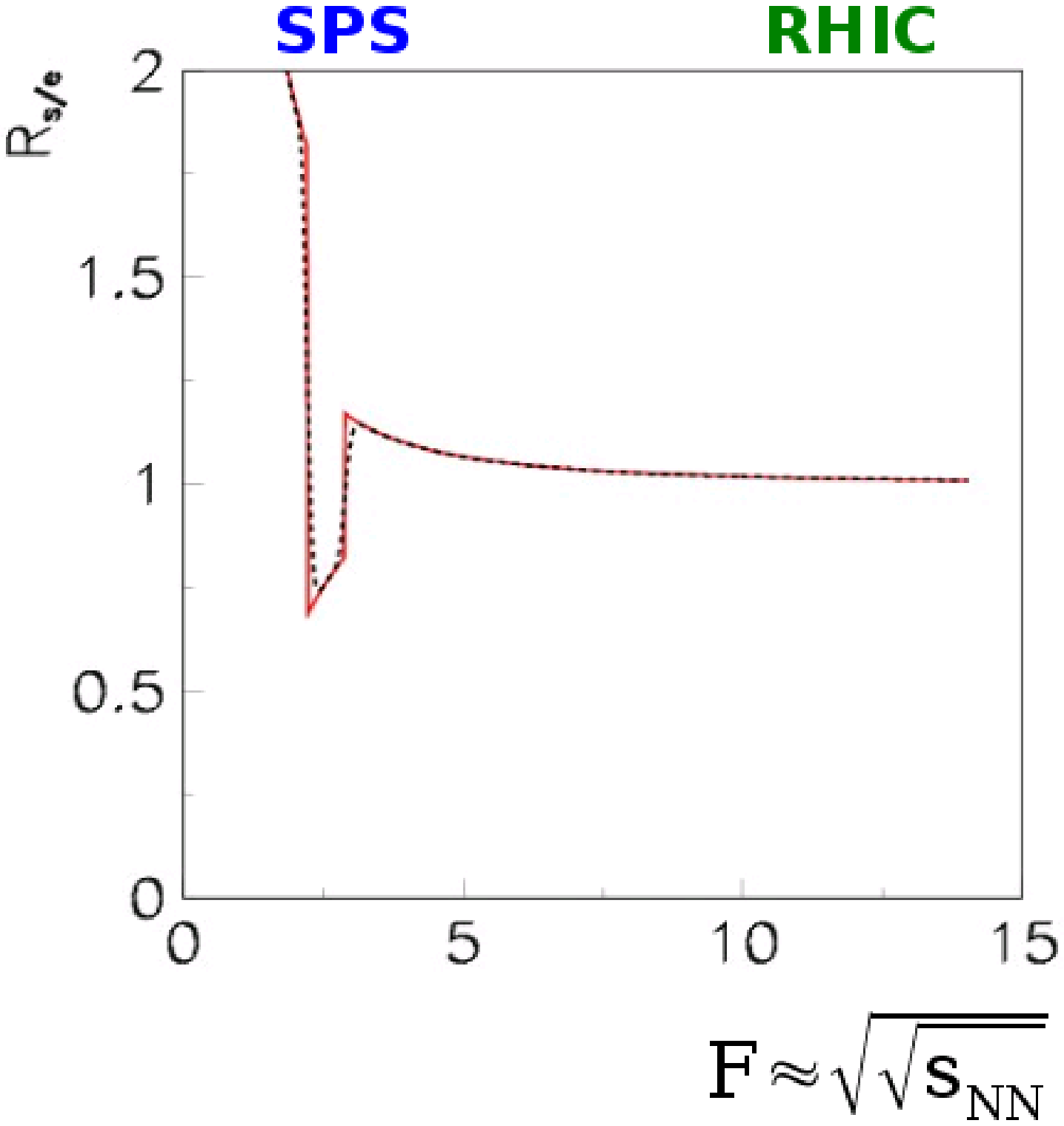}\hspace{1pc}%
\vspace{1pc}\caption{\label{sfluct}
The energy dependence ($F \equiv (\sqrt{s} - 2 m)^{3/4}/s^{1/8}$) 
of the properly filtered strangeness fluctuations,
$R_{s/e}$, calculated with SMES \protect\cite{gago,sfluct}.
The 'tooth' structure at the low SPS energies is caused by
by a change of fluctuations in the mixed phase region. 
}
\end{minipage}\hspace{1pc}%
\begin{minipage}{18pc}
\hspace{-1pc}\includegraphics[width=18pc]{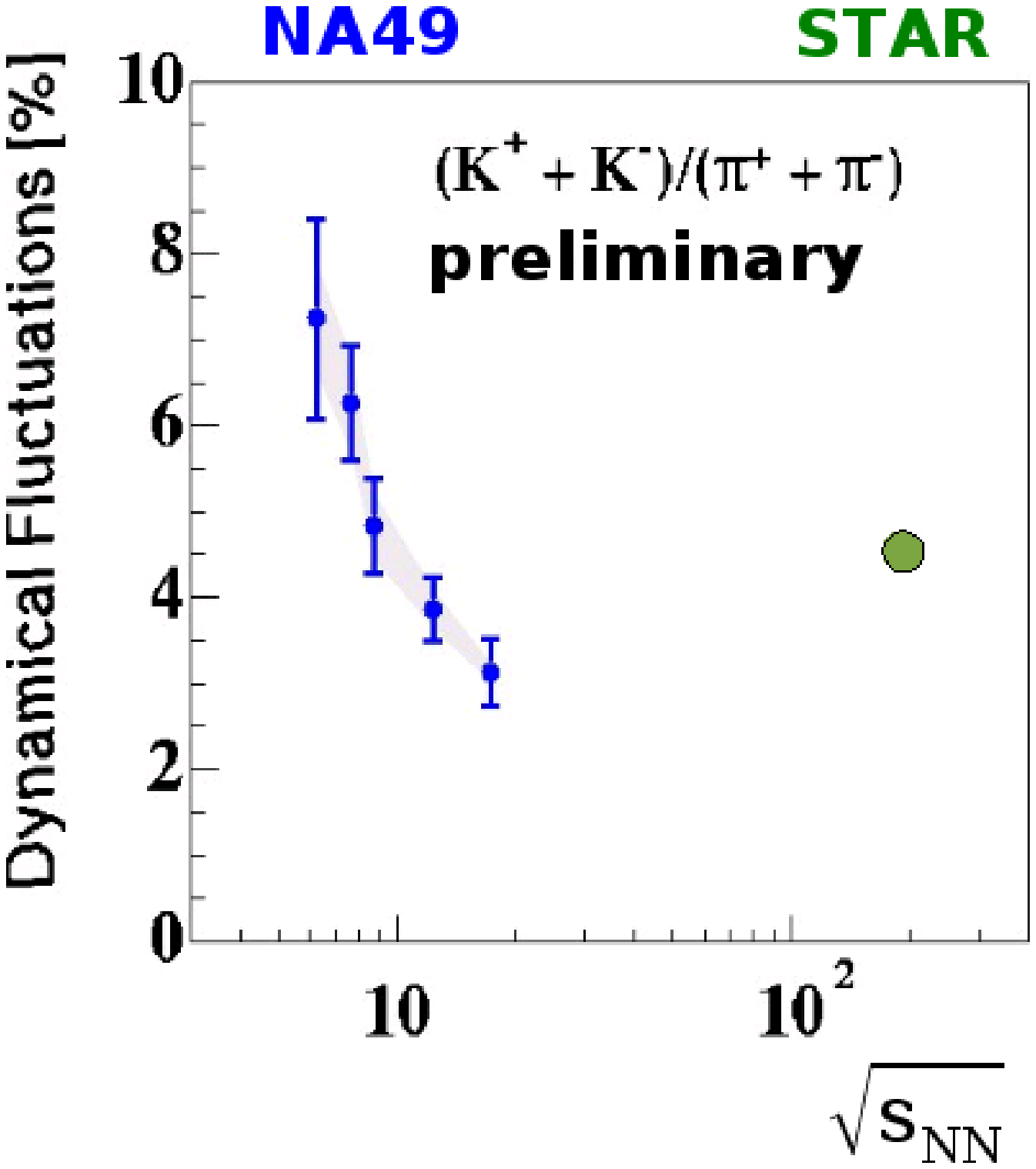}%
\caption{\label{sdata}
The energy dependence 
of the dynamical strangeness fluctuations in central
Pb+Pb collisions at the SPS energies \protect\cite{na49_s}
and central Au+Au collisions at the top RHIC energy 
\protect\cite{star_s}.
}
\end{minipage}
\end{figure}
The measure $R_{s/e}$ proposed to study them is defined as:
\begin{equation}
R_{e/s}  = \frac {(\delta \bar n_K)^2/\bar n_K^2} 
{(\delta \bar n)^2/ \bar n^2},
\end{equation}
where
$\delta \bar n_K$ and $\delta \bar n$ are dynamical
fluctuations of the multiplicity of kaons and pions, respectively.
The energy dependence of the measure $R_{s/e}$ calculated
within the SMES model \cite{gago, sfluct} is shown in 
Fig.~\ref{sfluct}. The observed rapid change of the behavior
is caused by the onset of deconfinement.
The first results of NA49 \cite{na49_s} and STAR \cite{star_s}
concerning strangeness fluctuations
in central Pb+Pb collisions at the SPS energies are  presented 
in Fig.~\ref{sdata}. The data seem to also 
suggest a change of energy dependence in  SPS energies.
However, a direct comparison between the data and the model
predictions is not possible as the used experimental measure
of fluctuations is  different than $R_{s/e}$.

\section{Summary}

The basic ideas concerning fluctuation signals of the
onset of deconfinement at the SPS energies were reviewed.
The status of their experimental tests based on the NA49
data was presented.

The naively expected significant suppression of the net-charge
fluctuations  is not observed. This may be due to dilution
of the initial charge fluctuations by resonance decays.

The study of energy dependence of multiplicity fluctuations
is in progress. It focuses on the search for the 
"shark fin" structure
expected at
the onset of deconfinement.
In this respect two unexpected observations were made recently.
Theory: the scaled variance of multiplicity distribution 
calculated within various statistical ensembles 
(micro-canonical, canonical and grand canonical) was found
to be different even in the thermodynamical limit.
Experiment: 
the scaled variance of multiplicity distribution for 
Pb+Pb collisions at 158$A$ GeV increases significantly when 
going from central to semi-peripheral collisions.

The preliminary results on the dynamical fluctuations of
the kaon to pion ratio in central Pb+Pb collisions indicate
a change of behavior at the SPS energies.
The relation of this interesting effect
to the onset of deconfinement is still unclear.
This is because a proper comparison between the data and the
model predictions is still missing.

\ack
I would like to thank the organizers of the Workshop on Correlations and
Fluctuations in Relativistic Nuclear Collisions (MIT, April 23-25, 2005) 
for the interesting and inspiring meeting.
This work was supported by the Virtual Institute VI-146
of Helmholtz Gemeinschaft, Germany.

\end{document}